\documentclass[reprint,prl,showpacs]{revtex4-1}
\usepackage{amsmath}
\usepackage{amsfonts}
\usepackage{bm}
\usepackage{hyperref}
\usepackage{graphicx}
\usepackage{subfigure}
\usepackage{color}

\usepackage{amsmath}
\usepackage{amsfonts}
\usepackage{bm}
\usepackage{hyperref}
\usepackage{graphicx}
\usepackage{subfigure}

\DeclareMathOperator{\sgn}{sgn} 

\newcommand{\bk}{{\bf k}}

\begin{document}
\title{Spin-orbit locking as a protection mechanism of the odd-parity superconducting state against disorder}

\author{Karen Michaeli and Liang Fu}
\affiliation{Department of Physics, Massachusetts Institute of Technology, Cambridge, MA 02139}

\begin{abstract}

Unconventional superconductors host a  plethora of  interesting physical phenomena. However, the standard theory of superconductivity suggests that unconventional pairing is highly sensitive to disorder, and hence can only be observed in ultraclean systems.   We find that due to an emergent chiral symmetry, spin-orbital locking can parametrically suppress pair decoherence introduced by impurity scattering  in odd-parity superconductors. 
Our work demonstrates that disorder is not an obstacle to realize odd-parity superconductivity in materials with strong spin-orbit coupling.  
\end{abstract}


\maketitle

% The reason for this protection is (a) only certain scatterings over the Fermi surface are harmful to superconductivity;  
%(b) for common types of disorder, the matrix elements of such scatterings are severely suppressed by the spin-orbital locked electron wavefunctions. 
%This novel mechanism 
%Specifically we study the effect of disorder on the odd-parity pairing in Cu$_x$Bi$_2$Se$_3$. 

%Moreover, the typical strength of spin-orbit coupling (defined below) is comparable to the Fermi energy\cite{hasan}. 
%We note that the wavefunctions $\psi_\alpha(\vec k)$ are not globally smooth over the Fermi surface. Instead there is an unavoidable  singularity at the south pole. As a result, both $\tilde{\Delta}$ and $V$ also acquire singularities.    Nonethess, we will show that physical quantities are non-singular.  

The quest for topological superconductors~\cite{ludwig, read, volovik, zhang, roy} is an
exciting research area in condensed matter physics. A
necessary condition for topological superconductors is
unconventional pairing symmetry. Based on an earlier
parity criterion~\cite{fukane}, it has been shown that under fairly
general conditions, time-reversal-invariant topological
superconductivity is realized if the pairing symmetry is
odd under spatial inversion~\cite{fuberg, sato}. Such odd-parity pairing
occurs in the Ballian-Werthamer phase of superfluid
helium-3~\cite{leggett}, and likely in certain heavy fermion superconductors~\cite{norman, sauls}.

A recent theoretical study~\cite{fuberg}  suggests that doped
narrow-gap semiconductors are candidates for odd-parity
topological superconductors. Here the strong spin-orbital
coupling in the band structure favors a novel interorbital,
odd-parity pairing, even when the mechanism for superconductivity
is conventional electron-phonon interaction.
Experimentally, several materials in this class, including
Cu-doped Bi$_2$Se$_3$~\cite{cava}, Tl-doped PbTe~\cite{pbte}, and In-doped
SnTe~\cite{snte}, exhibit superconductivity with unusually high
transition temperatures ($2-4$K) relative to their low carrier
densities ($\sim 10^{20}$cm$^{-3}$). Recently, some evidence of non
s-wave pairing in Cu$_x$Bi$_2$Se$_3$ has been reported~\cite{specific, magnetization, hc2},
including the presence of a zero-bias conductance peak in
point-contact spectroscopy~\cite{ando, point}.

An important issue in the study of unconventional superconductors is their robustness  
against disorder. Within the Bardeen-Cooper-Schrieffer  theory, s-wave pairing is immune to nonmagnetic impurities\cite{anderson}, while other pairing  symmetries are more fragile\cite{disorder}. For instance, the transition temperature of  Sr$_2$RuO$_4$ that is believed to be a spin-triplet superconductor is strongly suppressed by disorder\cite{sr2ruo4}. This seems to be a major obstacle for realizing odd-parity superconductivity in doped narrow-gap semiconductors.

In this Letter, we study the effect of disorder on the
proposed odd-parity superconducting state in narrow-gap
semiconductors~\cite{fuberg, hsiehfu}. We mostly focus on scalar impurities which are usually the most common type of disorder. Such an isotropic scattering potential arises
from Coulomb interactions between the electrons and
lattice defects. Contrary to conventional wisdom, we
find that the destructive pair-breaking effect of disorder
is dramatically suppressed by an approximate chiral
symmetry in the spin-orbital locked band structure. In
view of our study, the prospect of topological superconductivity
in narrow-gap semiconductors appears
brighter than before.

For concreteness, we start our analysis with the following non-interacting four band $\mathbf{k}\cdot\mathbf{p}$ Hamiltonian:
\begin{align} \label{H0}
\hspace{-1mm}H_{0}\hspace{-0.5mm}(\mathbf{k})\hspace{-1mm}=\hspace{-0.5mm}  \psi^\dagger({\bk})\hspace{-0.5mm} \left[\hspace{-0.25mm} m\hspace{-0.25mm}\sigma_x\hspace{-0.5mm} +\hspace{-0.5mm}v\sigma_z  \hspace{-0.25mm}(k_{x}\hspace{-0.25mm} s_y\hspace{-0.5mm} - \hspace{-0.5mm}k_{y}\hspace{-0.25mm} s_x\hspace{-0.25mm})\hspace{-0.5mm}+\hspace{-0.5mm} v_z\hspace{-0.25mm} k_{z} \hspace{-0.25mm}\sigma_y \hspace{-0.25mm}\right]\hspace{-0.5mm} \psi({\bk}). 
\end{align}
$H_0$ has the form of a massive Dirac Hamiltonian. Here, $v$ and $v_z$ are the velocities of the electrons in the $xÐy$ plane and $z$ direction, respectively, while the Dirac mass $m$ (not to be confused with the effective mass) determines the energy gap between the bands. Eq.~(\ref{H0})  describes the band structure of a broad class of narrow-gap semiconductors with inversion symmetry near time-reversal-invariant momenta~\cite{fukane}. For example, in the context of doped Bi$_2$Se$_3$, PbTe and SnTe, $s_i$ are Pauli matrices in spin space, and $\sigma_i$ are Pauli matrices associated with the orbital degrees of freedom~\cite{hsiehfu, hsieh}. Specifically for doped Bi$_2$Se$_3$, we use $\sigma_z=\pm1$ to label the two $p_z$-like orbitals located at the upper and lower part of the quintuple layer unit cell. The two orbitals ($+$ and $-$) transform into each other under spatial inversion with respect to the center of the unit cell.

\begin{figure}[bp]
\begin{flushright}\begin{minipage}{.5\textwidth}  \centering \subfigure[]{
        \label{fig:Sigma} %% label for first subfigure
        \includegraphics[width=1\textwidth]{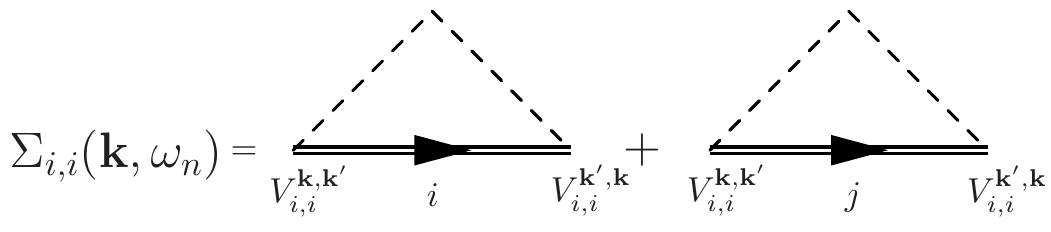}} \hspace{0.05in}%
        \subfigure[]{
        \label{fig:Cooperon} %% label for second subfigure
        \includegraphics[width=1\textwidth]{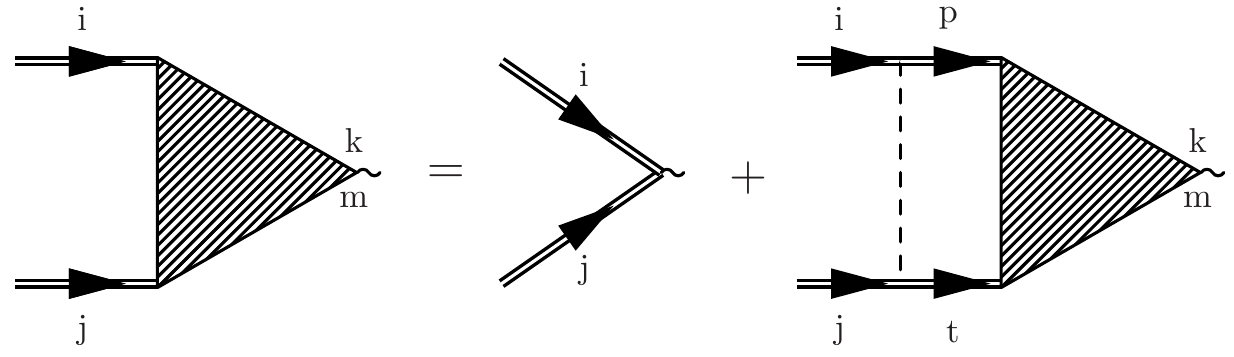}}
          \caption[0.4\textwidth]{\small The effect of disorder on the single particle Green's function and pair susceptibility. The self-consistent equation for the self-energy (Eq.~\ref{SelfEnergy}) is illustrated in (a). The double line represents the dressed electron Green's function, and the dashed line denotes the impurities.  The equation for the Cooperon (Eq.~\ref{Eq:Cooperon}) is illustrated in (b). Here, the  Cooperon is drawn as a filled triangle vertex.  }
\end{minipage}\end{flushright}
\end{figure}

 Short range (phonon mediated) attractive interactions can generate pairing with two distinct symmetries. One is the conventional spin-singlet 
intraorbital  state, while the other is the unconventional spin-triplet, orbital singlet paired state\cite{fuberg}:
\begin{subequations}
\begin{eqnarray}
\hspace{-10mm}\Delta_1\propto \hspace{-2mm}\sum_{\sigma=\pm}\hspace{-1mm}
\langle{\psi}_{\sigma,\uparrow}(\mathbf{k})\psi_{\sigma,\downarrow}(-\mathbf{k})-\psi_{\sigma,\downarrow}(\mathbf{k})\psi_{\sigma,\uparrow}(-\mathbf{k})\rangle;
\end{eqnarray} 
\begin{eqnarray}
\hspace{-5mm}\Delta_2 \propto \hspace{-2mm}\sum_{\sigma=\pm}\hspace{-1mm}\sigma\langle{\psi}_{\sigma,\uparrow}(\mathbf{k}) 
\psi_{-\sigma,\downarrow}(-\mathbf{k})+\psi_{\sigma,\downarrow}(\mathbf{k})\psi_{-\sigma,\uparrow}(-\mathbf{k})\rangle.
\end{eqnarray} 
\end{subequations}
The $\Delta_1$ pairing is invariant under all symmetry operations of the $D_{3d}$ point group ($s$-wave), whereas $\Delta_2$ is odd under spatial 
inversion (which interchanges the two orbitals) and belongs to the $A_{1u}$ representation. In Ref.~\onlinecite{fuberg} it has been shown that, in the absence of disorder, the strength of the interorbital and intraorbital attractive interaction determines which pairing susceptibility diverges more strongly. Here, we examine the effect of disorder on the odd-parity pairing assuming the interaction favors this state. In contrast to the well known $p$-wave spin-triplet superconductivity, here the $\Delta_2$ order parameter is independent of  momentum, but has a nontrivial internal structure in orbital and spin space. In the traditional $p$-wave superconductors the  order parameter varies over the Fermi surface, and scattering of a Cooper pair from the states $\mathbf{k},-\mathbf{k}$ into  $\mathbf{k}',-\mathbf{k}'$ results in their decoherence. Consequently, superconductivity dies when  the elastic scattering rate becomes comparable to the order parameter, $1/\tau\rightarrow\Delta$. Our result shows that  decoherence effects   in the spin-triplet, orbital-singlet paired state are significantly suppressed, and  superconductivity survives much stronger levels of disorder until $~(m/\mu)^2/\tau\rightarrow\Delta$.

To  show quantitatively   the  effect of disorder on the onset of superconductivity, we examine the pairing susceptibility near the transition.  
First, we transform into the eigenstate basis, in which  the Hamiltonian is diagonal,  $H_0=\sum_{\bk, j} E(\bk) \left[ c^\dagger_{j}(\bk) c_{j}(\bk) -  d^\dagger_{j}(\bk) d_{j}(\bk)\right]$, where  the   dispersion  is $E(\mathbf{k}) =\sqrt{m^2+ v^2 (k_x^2+ k_y^2)+v_z^2k_z^2}$. Despite the strong spin-orbital mixing in $H_0$,  the presence of both time-reversal and inversion symmetries protects the two-fold degeneracy of the upper ($c_1$, $c_2$) and lower  ($d_1$, $d_2$)  bands at every $\bk$. We chose to work in a particular basis~\cite{Supp},  which we  dub  $"$pseudo-chiral$"$, where the states at $\bk$ and $-\bk$ are related (up to a $\bk$-dependent phase factor) by time reversal ($\Theta$) and inversion ($P$) operations:
\begin{eqnarray}
\Theta  c_1 (\bk) \Theta^{-1} \sim c_1(-\bk)&,& \; \Theta  c_2 (\bk) \Theta^{-1} \sim c_2(-\bk)   \nonumber \\
P c_1(\bk) P^{-1} \sim c_2(-\bk)&,&  \; P c_2(\bk) P^{-1} \sim c_1(-\bk). \label{sym}
\end{eqnarray}
As long as the Hamiltonian has both time-reversal  and inversion symmetries, 
it is always possible to use the basis satisfying  the transformation properties described in Eq.~\ref{sym}. 
%Thus, with this general definition, our theory  applies to disorder effects on odd-parity superconductivity in generic spin-orbit coupled systems. 

In the pseudo-chiral basis, the superconducting order parameters (2a) and (2b) are given by:
\begin{subequations}\label{delta}
\begin{align}\label{delta1}
\Delta_{1}(\bk)\hspace{-0.5mm} &\propto\hspace{-0.5mm} {e}^{-i\phi}  \sum_{j=1,2} (-1)^j \left[ \langle{c_{j}(\bk)c_{j}(-\bk) \rangle + \langle d_j(\bk)  d_j(-\bk) }\rangle \right]
\end{align}
\begin{align}\label{delta2}
&\Delta_{2}(\bk) \hspace{-0.5mm} \propto\hspace{-0.5mm} {e}^{-i\phi}\cos\alpha_\bk \hspace{-1.5mm} \sum_{j=1,2} \hspace{-1mm} \left[ \langle c_{j}(\bk)c_{j}(-\bk) \rangle  \hspace{-0.5mm}- \hspace{-0.5mm} \langle d_j(\bk)  d_j(-\bk) \rangle \right] \nonumber\\
&- 2{e}^{-i\phi}\sin\alpha_\bk \left[ \langle c_1(\bk)d_2(-\bk) \rangle+ \langle c_{2}(\bk) d_1(-\bk) \rangle  \right].
\end{align}
\end{subequations}
In the derivation of the above expressions and from now on we rescaled the $z$ component of the momentum  by its velocities, $\mathbf{k}=(k_x,k_y,v_zk_z/v)$. While the expressions for the order parameters contain the azimuthal angle  $\phi$  between $(k_x, k_y)$ and the $x$ axis, they are independent of the polar angle $\theta$ between $\mathbf{k}$ and the $z$-axis. The parameter $\alpha_\bk=\sin^{-1}(m/E(\bk))$  is a consequence of the interorbital mixing by the max term.

In the limit $m=0$ ($\alpha_{\mathbf{k}}=0$), the $\mathbf{k}\cdot\mathbf{p}$ Hamiltonian has a $U(1)$ chiral symmetry, $[H_0,\sigma_ys_z]=0$. Thus, the energy eigenstates labeled by $j=1(2)$ have a well-defined chirality $+1(-1)$, which is evident from Eq.~\ref{delta}. Importantly,  pairing in both the even- and odd-parity states only occur between electrons of the same chirality, but the order parameters differ by the relative phase between Cooper pairs of opposite chirality (of different label $j$), $\pi$ for $\Delta_1$ and $0$ for $\Delta_2$\cite{fuberg}.  Since  scalar disorder potential (which is insensitive to orbitals and spins) does not break  chiral symmetry,  impurities can scatter electrons only between states  of equal chirality. Hence, the disorder affects the  even-parity pairing $\Delta_1$ and odd-parity pairing $\Delta_2$ in the same way. Then, it follows from the Anderson theorem~\cite{anderson} that as long as the system is far from localization, both pairings are completely robust against disorder. The magnitudes of two order parameters at zero temperature differ only by the strength of the pairing interaction in the two channels, $\Delta_{\ell=1,2}=\Omega\exp\{-1/\nu_0\lambda_{\ell}\}$  with $\nu_0$ the density of states at the Fermi energy, $\lambda_{\ell}$ the attractive interaction in the even- or odd-parity channel, and  $\Omega$ an ultraviolet cutoff. 

For nonzero $m$ in the Hamiltonian $H_0$,  the  Bloch wave-functions are  no longer chiral eigenstates, and the similarity between the two order parameters is  broken. Without  loss of generality, we assume that the chemical potential  lies in the upper energy  band,  $\mu>|m|$.  Since only electrons on the Fermi surface contribute to the divergent pairing susceptibility at low temperature, 
 the {\it effective} pairing order parameter involves only the $c$ electrons of the conduction bands:
\begin{subequations} \label{pairing}
\begin{eqnarray}
\hspace{-15mm} \Delta_{1}(\bk) &\propto& {e}^{i\phi}   \left[  \langle  c_{1}(\bk)c_{1}(-\bk) \rangle  - \langle  c_{2}(\bk)c_{2}(-\bk) \rangle  \right] 
\end{eqnarray}
\begin{eqnarray}
\hspace{-5mm}\Delta_{2}(\bk) &\propto& {e}^{i\phi}\cos\alpha \left[  \langle  c_{1}(\bk)c_{1}(-\bk) \rangle  + \langle  c_{2}(\bk)c_{2}(-\bk) \rangle  \right]. \label{pairing}
\end{eqnarray} 
\end{subequations}
Here $\alpha = \sin^{-1}(m/\mu)$ is evaluated at the Fermi energy. One can see that for $m\neq0$ the effective pairing potential  in the odd-parity channel between electrons in the conduction bands is reduced due to mixing with the valence electrons   (see Eq.~\ref{delta2}). Assuming the strength of the inter-orbital attraction is independent of  the parameter $m/\mu$, the magnitude of the odd-pairing order parameter at zero temperature  becomes  $\Delta_{2}=\Omega\exp\{-1/\nu_{\alpha}\cos^2\alpha\lambda_{\ell}\}$.   The order parameter  $\Delta_2$ is reduced from its $m=0$ value not only due to the  dependence of the density of states  on $m/\mu$, $\nu_{\alpha}=\nu_0\cos\alpha$, but mainly because the attractive interaction is weaker by a factor of $\cos\alpha$. We wish to emphasize that this change in the odd-symmetry order parameter is not caused by pair breaking. Pair breaking effects  reduce the transition temperature, but not the zero temperature order parameter\cite{Shiba}.

Introducing scattering by a scalar disorder potential  adds a term into the Hamiltonian which is non-diagonal in the momentum, $H(\mathbf{k},\mathbf{k}')=H_{0}(\mathbf{k})\delta_{\mathbf{k},\mathbf{k}'}+V_{\text{imp}}\psi^{\dag}(\mathbf{k})\psi(\mathbf{k})$. Transforming into the pseudo-chiral basis, one can see that the impurities mix all bands. However, in the limit  $\varepsilon_F\tau\gg1$, all leading order processes occur near the Fermi surface, and we can again restrict our attention to the conduction bands. The matrix element between two states at the Fermi energy, $\psi_{i}(\mathbf{k})$ and $\psi_{j}(\mathbf{k}')$, is given by
\begin{align}
V_{\mathbf{k}, \mathbf{k'}}^{i,j} &=  
\begin{pmatrix}
    A(\mathbf{k}, \mathbf{k}')  &  A(\mathbf{k}, -\mathbf{k}') \sin \alpha \\
   A(-\mathbf{k}, \mathbf{k}') \sin \alpha    &    A(-\mathbf{k}, -\mathbf{k}') 
    \end{pmatrix}
\end{align}
Here $A(\mathbf{k}, \mathbf{k}')=V_{\text{imp}}[e^{i(\phi-\phi')}\cos\frac{\theta}{2}\cos\frac{\theta'}{2}+\sin\frac{\theta}{2}\sin\frac{\theta'}{2}]$ is 
equal to the wavefunction overlap between two spins pointing along $\bk$ and $\bk'$ directions. In the limit $m=0$ ($\alpha=0$) the off-diagonal matrix elements vanish, restoring the chiral symmetry. 

To find the corrections to the electron self-energy $\Sigma$ due to scattering by disorder we use the self-consistent Born approximation. Then the self-energy matrix for the two conduction bands is diagonal and determined by two processes, intra-band scattering and  scattering between the two conduction bands (for illustration see Fig.~\ref{fig:Sigma}). The  self energy is found from the following self-consistent equation:
\begin{align}\label{SelfEnergy}\nonumber
\Sigma_{\mathbf{k},\omega_n}&\equiv-\frac{{i}}{2\tau}\sgn(\omega)=\int\frac{d\mathbf{k}'}{(2\pi)^3} |V_{\mathbf{k}, \mathbf{k'}}^{i,i}|^2G_{\mathbf{k}',\omega_n}^{i}\\
&+\int\frac{d\mathbf{k}'}{(2\pi)^3} V_{\mathbf{k}, \mathbf{k'}}^{i,j\neq{i}}V_{\mathbf{k}', \mathbf{k}}^{j\neq{i},i}G_{\mathbf{k}',\omega_n}^j,
\end{align}
where $G_{i}^{-1}(\mathbf{k},\omega_n)=i\omega_n-(E(\mathbf{k})-\mu)-\Sigma(\mathbf{k},\omega_n)$ is the Matsubara Green's function of the electrons in the band $i$. Examining the above equation,  one can see that the two scattering processes interfere constructively. Consequently, the elastic scattering time satisfies $1=\pi\nu_{\alpha}{V}_{imp}^2\tau(1+\sin^2\alpha)$.

As we explained above, while the transition temperature is highly sensitive to  pair decoherence mechanisms, the zero temperature order parameter is not  modified. Therefore, we calculate  the transition temperature   into the superconducting phase for the even- and odd-pairing states from the corresponding  pairing susceptibilities of the normal state:
\begin{align}
\chi^{\ell}=\lambda{T}\sum_{ \begin{matrix}
  \vspace{-1.5mm} \scriptstyle{i,j,k,m}  \\  \scriptstyle{\mathbf{k},\omega_n}
   \end{matrix}} \Gamma_{i,j}^{\ell}G_{\mathbf{k},\omega_n}^{i}G_{-\mathbf{k},-\omega_n}^{j}C_{i,j;k,m}(\omega_n)\Gamma_{k,m}^{\ell}.
\end{align} 
Here, $\Gamma_{i,j}^{\ell}=\delta_{i,j}[\delta_{i,1}-\delta_{i,2}]$  for the spin-singlet pairing $\ell=1$, and $\Gamma_{i,j}^{\ell}=\delta_{i,j}\cos\alpha[\delta_{i,1}+\delta_{i,2}]$  for the orbital-singlet pairing $\ell=2$. The matrix $C_{i,j;k,m}(\omega_n)$ is the Cooperon describing multiple scattering events of two electrons (a Cooper pair) in the particle-particle channel.  A pole in the  Cooperon  at $\omega_n=0$  means that scattering by impurities does not result in  decoherence of  Cooper pairs. In other words, the probability of an electron in state $\mathbf{k}$ to be scattered into state $\mathbf{k}'$ is equal to the probability of its partner
to be scattered into the partner of $\mathbf{k}'$. A Cooperon with a finite mass, on the other hand, implies that  pairing is  suppressed by disorder and that in the superconducting state there are sub-gap excitations.  

The Cooperon can be expressed in terms of the single particle Green's function and impurity scattering potential:
\begin{align}\label{Eq:Cooperon}
&C_{i,j;k,m}(\omega_n)=
\delta_{i,k}\delta_{j,m}+\\\nonumber&\sum_{p,t=1,2}
\int\frac{d\mathbf{k}'}{(2\pi)^3}V_{\mathbf{k},\mathbf{k}'}^{i,p}V_{-\mathbf{k},-\mathbf{k}'}^{j,t}G_{\mathbf{k}',\omega_n}^{p}G_{-\mathbf{k}',-\omega_n}^tC_{p,t;k,m}(\omega_n).
\end{align} 
The only four components of the Cooperon that enter the susceptibilities are  $C_{i,i;j,j}$. In the absence of impurities  $C_{1,1;2,2}=C_{2,2;1,1}=0$ and only the two  components $C_{1,1;1,1}=C_{2,2;2,2}$ matter.  As a result, there is no difference in the effect of disorder on both order parameters.  

Similar to the single-particle  self-energy,  the Cooperon includes processes in which the electrons remain in the same band after scattering and those in which at least one electron changes its band. However, while the interference between  the scattering events that determine the elastic scattering time is constructive, the Cooperon is a sum of constructive and destructive interference terms. This can be seen in the expressions for  $C_{i,i;j,j}$:
\begin{align}
C_{i,i;j,j}&=\frac{1+2|\omega_n|\tau}{1+2|\omega_n|\tau-\pi\nu_{\alpha}{V}_{imp}^2\tau(1-\sin^2\alpha)}
\\\nonumber&+(-1)^{i+j}\frac{1+2|\omega_n|\tau}{1+2|\omega_n|\tau-\pi\nu_{\alpha}{V}_{imp}^2\tau(1+\sin^2\alpha)}.
\end{align} 
Note that the above expression for the Cooperon has been calculated only assuming that $\varepsilon_F\tau\gg1$.   Now the two susceptibilities, $\chi^{1}$ and $\chi^{2}$, are no longer identical. The even-pairing susceptibility includes only the constructive interference processes, $\chi^{1}=4\pi{T}\sum_{\omega_n}\nu_{\alpha}\tau[1-\pi\nu_{\alpha}{V}_{imp}^2\tau(1+\sin^2\alpha)(1-2|\omega_n|\tau)]^{-1}=2\pi{T}\sum_{\omega_n}\nu_{\alpha}/|\omega_n|$, and is clearly insensitive to  disorder. In contrast, the susceptibility for odd-pairing is determined by the destructive interference, and hence, affected by impurities: 
\begin{align}\label{Susc}
&\chi^{2}=T\sum_{\omega_n}\frac{\pi\nu_{\alpha}\cos^2\alpha}{|\omega_n|+{\sin^2\alpha}/{\tau(1+\sin^2\alpha)}}\\\nonumber&=2\nu_{\alpha}(\cos^2\alpha)\left[\ln\frac{\Omega}{2\pi{T}}-\psi\left(\frac{1}{2}+\frac{\sin^2\alpha}{2\pi{T}(1+\sin^2\alpha)\tau}\right)\right],
\end{align} 
where $\psi(x)$  is the digamma function.

\begin{figure}[tp]
\begin{flushright}\begin{minipage}{0.5\textwidth}  \centering
        \includegraphics[width=1\textwidth]{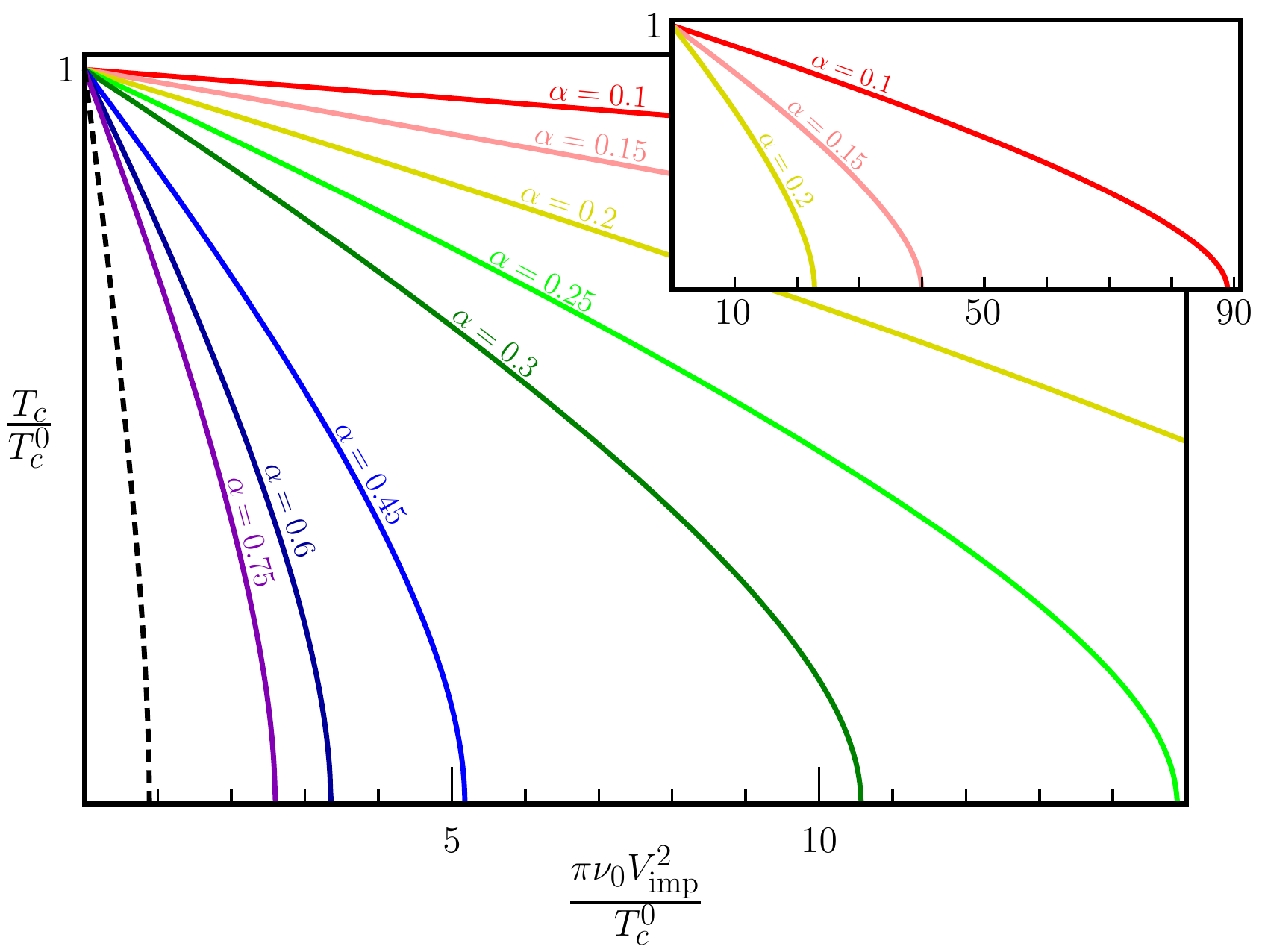} \hspace{0.05in}
                 \caption[0.4\textwidth]{\small The critical temperature as a function of the level of disorder $\pi\nu_0V_{\text{imp}}^2$ for  various values of $\alpha%=0.1,0.15,0.2,0.25,0.3,0.45,0.6,0.75
                 $. Both $T_c$ and $\pi\nu_0V_{\text{imp}}^2$ are given in units of 
of $T_c^0$, the critical temperature for $V_{\text{imp}}=0$. For comparison, the dashed line shows the transition temperature into the odd-parity superconducting state in the absence of spin-orbital locking (as in Helium-III). One can see that up to $\alpha\approx0.5$ the critical level of disorder in which the unconventional order parameter disappears is dramatically higher than in Helium-III. At $\alpha<0.3$ the effect of moderate disorder on $T_c$ is almost unnoticeable. 
%Since $\alpha$ can be tuned by changing the chemical potential, we believe that odd-parity superconductivity can be easily observed in doped narrow-gap semiconductors. 
}
                 \label{fig:diagrams}
\end{minipage}\end{flushright}
\end{figure}

Let us explain the above result. The pair decoherence rate, $1/\tau_{\varphi}=(\sin^2\alpha)/(1+\sin^2\alpha)\tau$, which reduces the transition temperature enters  the Cooperon as a mass term eliminating its divergence. When $1/\tau_{\varphi}$ is larger than the order parameter in the absence of pair decoherence, superconductivity  disappears.  We found that disorder introduces a  pair breaking mechanism for the odd-parity pairing but not for the even-parity pairing, and that the only difference between the two order parameters is the phase between the condensates in the two conducting bands. Therefore, we conclude that the decoherence is due to the tendency of the impurity scattering  to  favor a relative phase of $\pi$ between the conduction bands (the singlet pairing), suppressing the spin-triplet transition temperature:
\begin{align}\label{Tc}
\ln\frac{T_c(\tau)}{T_c^0}&=
\psi\left(\frac{1}{2}\right)-\psi\left(\frac{1}{2}+\frac{1}{2\pi{T}_c(\tau)\tau_{\varphi}}\right).
\end{align} 
Here we use $T_c^0$  to denote the transition temperature  at $1/\tau=0$ and finite $\alpha$, and $C=\pi/2e^{\gamma}$ with $\gamma$ the Euler constant.  The suppression of $T_c$ as a function of disorder for various values of $\alpha$ is illustrated in Fig.~\ref{fig:diagrams}.

To understand better the peculiarity of this result, it is instructive to compare our result, applicable for the narrow-gap semiconductors, with the BW superfluid phase in He-III.  Although the latter has a single band without  spin-orbit coupling, one can choose a  basis 
in which spin is locked to be parallel, $\phi_1(\mathbf{k})$, or anti-parallel, $\phi_2(\mathbf{k})$ to the 
momentum. This basis $\phi $ resembles the chiral basis we used here.  Correspondingly, in both cases the odd-parity order parameters  can  be 
written as $e^{i\phi}[\phi_1(\mathbf{k}) \phi_1(-\mathbf{k}) + \phi_2(\mathbf{ k})\phi_2(-\mathbf{k})]$. The key difference between systems with and without spin-orbital locking manifests itself in the impurity scattering.  While in Helium-III , the matrix element for impurity scattering between $\phi_1(\mathbf{k})$ and  $\phi_2(\mathbf{k}')$ are larger than the diagonal terms, in the problem studied here the inter-band scattering is parametrically smaller than the intra-band one  by a factor of $\sin \alpha = m/\mu$, due to the approximate chiral symmetry that becomes exact as $m \rightarrow 0$. Thus,  the pair decoherence in the narrow-gap semiconductors is significantly weaker than in Helium-III. 
Note that this observation is only correct for scalar disorder, and does not hold for other types of scattering potentials, such as magnetic impurities ($\propto\vec{s}$)  or orbital dependent potentials   ( $\propto\vec{\sigma}$). These non-scalar disorder potentials cause stronger pair decoherence and suppresses the unconventional superconducting state.  %We chose to concentrate here on the effect of a scalar impurity potential, as it is usually the dominant type of disorder.

Our result can be generalized to 2D bilayer band structures obtained by setting $k_z=0$ in the $k\cdot p$ Hamiltonian~(\ref{H0}). Recently, it has been proposed~\cite{nagaosa} that this class of bilayer systems with Rashba spin-orbit coupling favors odd-parity superconductivity similar to the 3D case. Our analysis shows a similar robustness against disorder due to
the chiral symmetry (See Supplemental Material~\cite{SM}).

To conclude, we showed that certain types of odd-parity pairing in doped narrow-gap semiconductors can survive from a fairly large amount of impurity scattering. The relative robustness of these systems results from an approximate chiral symmetry in the spin-orbital locked band structure and the odd-parity pairing order parameter. Although scattering by disorder reduces the phase coherence of the Cooper pairs, the dephasing rate  $1/\tau_{\varphi}=\pi\nu_{\alpha}V_{\text{imp}}^2\sin^2\alpha$ vanishes as the Dirac mass (= band gap) in the band structure goes to zero, when the chiral symmetry becomes exact. Finally, we note that in addition to centrosymmetric materials studied in this work, strong spin-orbit-coupling in asymmetric interface structures can also protect unconventional superconductivity against
disorder~\cite{michaeli}.

{\it Acknowledgement:} This work is supported by Pappalardo Fellowship (KM) and start-up funds from MIT (LF). We thank 
Dung-hai Lee for motivating our interest in the effect of disorder in odd-parity superconductors and Yang Qi for helpful discussions.

\newpage

\section{Supplementary materials}

\subsection{Choice of basis}
In the supplementary material part, we elaborate on the transformation connecting between the basis of spin ($s=\uparrow, \downarrow$) and orbital ($\sigma=\pm$) eigenstates and the pseudo-chiral basis. The creation operators of electrons in the conduction bands, $c_j^{\dag}$, and valence bands, $d_j^{\dag}$  can be written as:     
\begin{align}
c^\dagger_j(\bk) &=& \sum_{\sigma, s} u_{\bk, j}(\sigma, s) \psi_{\sigma,s}^\dagger(\bk)   , \nonumber \\
d^\dagger_j(\bk) &=& \sum_{\sigma, s} v_{\bk, j}(\sigma, s) \psi_{\sigma,s}^\dagger(\bk).
\end{align}
The functions $u_{\bk, j}$ and $v_{j}$ are four components vectors in the spin and orbital space, $(+\uparrow, +\downarrow, -\uparrow, -\downarrow)$. 
For convenience,  we rescale $k_z$ by its  velocity, $\mathbf{k}=(k_x,k_y,v_zk_z/v)$. The transformation functions $u_{\bk,j}$ to the conduction bands are  given by 
\begin{align}
u_{\bk, 1}&=& \frac{1}{\sqrt{2}}( i\cos\frac{\theta}{2} e^{-i \alpha/2},  \sin \frac{\theta}{2} e^{i (\phi + \alpha/2)} , \nonumber \\
&&  \cos \frac{\theta}{2} e^{i \alpha/2}, i\sin \frac{\theta}{2} e^{i (\phi - \alpha/2)} ), \nonumber \\
u_{\bk, 2}&=& \frac{1}{\sqrt{2}}( \sin\frac{\theta}{2} e^{-i \alpha/2}, i \cos \frac{\theta}{2} e^{-i (\phi - \alpha/2)} , \nonumber \\
&&   -i\sin \frac{\theta}{2} e^{i \alpha/2}, - \cos \frac{\theta}{2} e^{-i (\phi + \alpha/2)} ).
\end{align}
where $\theta$ and $\phi$ are the polar and azimuth angle of $\bk$, respectively, and $\alpha =\sin^{-1}(m/\mu)$. 
Similarly, for the valence bands:
\begin{align}
v_{\bk, 1}&=& \frac{1}{\sqrt{2}}( -\cos\frac{\theta}{2} e^{-i \alpha/2}, -i \sin \frac{\theta}{2} e^{i (\phi + \alpha/2)} , \nonumber \\
&&  -i\cos \frac{\theta}{2} e^{i \alpha/2}, -\sin \frac{\theta}{2} e^{i (\phi - \alpha/2)} ), \nonumber \\
v_{\bk, 2}&=& \frac{1}{\sqrt{2}}(i \sin\frac{\theta}{2} e^{-i \alpha/2},  \cos \frac{\theta}{2} e^{-i (\phi - \alpha/2)} , \nonumber \\
&&   -\sin \frac{\theta}{2} e^{i \alpha/2}, -i \cos \frac{\theta}{2} e^{-i (\phi + \alpha/2)} ).
\end{align}

Using the above transformation we obtained the expression for the Hamiltonian, order parameters and the impurity scattering matrix in the pseudo-chiral basis. This transformation allowed us to focus on  pairing, as well as elastic scattering events in the vicinity of the Fermi surface. 

\subsection{2D bilayer Rashba systems}

In this section we discuss the effect of disorder on unconventional pairing in 2D bilayer systems with Rashba spin-orbit coupling. Such a system is described by the the 3D Hamiltonian of Eq.~(1) in the main text with  $k_z=0$:
\begin{align}
H_{\rm 2D}(\mathbf{k})= \psi^\dagger({\bk}) \left[m\sigma_x +v\sigma_z  (k_{x}s_y -k_{y}s_x)\right] \psi({\bk}).
\end{align}
Here $\sigma_z=\pm 1$ is the layer index and $s_z=\pm 1$ is the electron spin along the direction perpendicular to the planes.   
As explained in Ref.~\cite{hsiehfu},  the top layer is subject to an electric field created by the bottom layer, and vice versa. These fields  lead to an intra-layer Rashba spin-orbit coupling term with an opposite sign for the two layers $\sigma_z  (k_{x}s_y -k_{y}s_x)$.  In the 2D Hamiltonian,  the mass term generates inter-layer hopping. The above Hamiltonian applies to a wide class of materials including thin films/quantum wells of narrow-gap semiconductors like Bi$_2$Se$_3$, as well as oxide heterostructures\cite{nagaosa}. 

The possible pairing symmetries coincide with those suggested in Ref.~\cite{fuberg} for the 3D Hamiltonian~\cite{nagaosa}. In particular, for $k_z=0$ the orbital-singlet spin-triplet pairing ($\Delta_2$) studied in the main text,  corresponds to inter-layer spin-triplet pairing in the bilayer system. Due to the same chiral symmetry discussed in the main text, there is no difference in the effect of disorder on $\Delta_2$ in the 2D and 3D cases. 

In addition to the two pairing symmetries discussed in the main text, Ref.~\cite{fuberg} proposed the following unconventional spin-singlet state:
\begin{align}
\hspace{-10mm}\Delta_3\propto \hspace{-3mm}\sum_{\mathbf{k},\sigma=\pm}\sigma
\langle{\psi}_{\sigma,\uparrow}(\mathbf{k})\psi_{\sigma,\downarrow}(-\mathbf{k})-\psi_{\sigma,\downarrow}(\mathbf{k})\psi_{\sigma,\uparrow}(-\mathbf{k})\rangle;
\end{align} 
Unlike the case of $\Delta_2$, the effect of disorder on this paired state dramatically changes when dimensionality is reduced from three to two dimensions. In the 3D case, $\Delta_3$ has nodes,  and the paired state is not protected by the chiral symmetry even when $m=0$. As a result, pairing is strongly suppressed by disorder. In contrast, in 2D the order parameter, which corresponds to spin-singlet 2D topological superconductor\cite{nagaosa} with opposite phase on the two layers, is fully gapped. Moreover, in the limit   $m \rightarrow 0$, a new chiral symmetry emerges and the Hamiltonian is invariant under  the transformation  $\psi_{\sigma,s} \rightarrow e^{i\sigma\theta}\psi_{\sigma,s}$. For $\theta=\pi/4$,  the odd-parity order parameter $\Delta_3$ transforms into  $\Delta_1$.  Thus, as long as this symmetry exists,  $\Delta_3$ is protected  against the effect of  both scalar and $\sigma_z$ type of impurities. For a finite mass term the suppression of  the transition temperature into $\Delta_3$  is parameterically weaker than it is in the 3D case.   It will be interesting to find such a $s_\pm$ odd-parity topological superconductor.

\end{thebibliography}

\end{document}